\documentclass[sigconf,natbib=false]{acmart}

\AtBeginDocument{%
  }
\copyrightyear{2025}
\acmYear{2025}
\setcopyright{acmlicensed}
\acmConference[WWW Companion '25]{Companion Proceedings of the ACM Web Conference 2025}{April 28-May 2, 2025}{Sydney, NSW, Australia}
\acmBooktitle{Companion Proceedings of the ACM Web Conference 2025 (WWW Companion '25), April 28-May 2, 2025, Sydney, NSW, Australia}
\acmDOI{10.1145/3701716.3715243}
\acmISBN{979-8-4007-1331-6/25/04}

\RequirePackage[
  datamodel=acmdatamodel,
  style=acmnumeric
  ]{biblatex}

\usepackage{afterpage}
\usepackage{multicol}
\begin{document}

%%
%% The "title" command has an optional parameter,
%% allowing the author to define a "short title" to be used in page headers.
\title{Latent Customer Segmentation and Value-Based Recommendation Leveraging a Two-Stage Model with Missing Labels }

%%
%% The "author" command and its associated commands are used to define
%% the authors and their affiliations.
%% Of note is the shared affiliation of the first two authors, and the
%% "authornote" and "authornotemark" commands
%% used to denote shared contribution to the research.
\author{Keerthi Gopalakrishnan}
\authornote{Both authors contributed equally to this work.}
\affiliation{%
  \institution{Walmart Global Tech}
  \city{Sunnyvale}
  \state{California}
  \country{USA}
  %\thanksref{equal}
}
\email{k.goppalakrishnan@walmart.com}

\author{Tianning Dong}
\authornotemark[1]
\affiliation{%
  \institution{Walmart Global Tech}
  \city{Sunnyvale}
  \state{California}
  \country{USA}
  %\thanksref{equal}
}
\email{Tianning.Dong@walmart.com}

\author{Chia-Yen Ho}
\affiliation{%
  \institution{Walmart Global Tech}
  \city{Sunnyvale}
  \state{California}
  \country{USA}
}
\email{chiayen.ho@walmart.com}

\author{Yokila Arora}
\affiliation{%
  \institution{Walmart Global Tech}
  \city{Sunnyvale}
  \state{California}
  \country{USA}
}
\email{yokila.arora@walmart.com}

\author{Topojoy Biswas}
\affiliation{%
  \institution{Walmart Global Tech}
  \city{Sunnyvale}
  \state{California}
  \country{USA}
}
\email{topojoy.biswas@walmart.com}

\author{Jason Cho}
\affiliation{%
  \institution{Walmart Global Tech}
  \city{Sunnyvale}
  \state{California}
  \country{USA}
}
\email{jason.cho@walmart.com}

\author{Sushant Kumar}
\affiliation{%
  \institution{Walmart Global Tech}
  \city{Sunnyvale}
  \state{California}
  \country{USA}
}
\email{sushant.kumar@walmart.com}

\author{Kannan Achan}
\affiliation{%
  \institution{Walmart Global Tech}
  \city{Sunnyvale}
  \state{California}
  \country{USA}
}
\email{kannan.achan@walmart.com}
%\thanks[equal]{These two authors contributed equally to this work.}

%%
%% By default, the full list of authors will be used in the page
%% headers. Often, this list is too long, and will overlap
%% other information printed in the page headers. This command allows
%% the author to define a more concise list
%% of authors' names for this purpose.
\renewcommand{\shortauthors}{Keerthi Gopalakrishnan et al.}
\renewcommand{\shorttitle}{Latent Customer Segmentation and Value-Based Recommendation}

%%
%% The abstract is a short summary of the work to be presented in the
%% article.
\begin{abstract}
\footnote{This is the author's version of the work. The definitive published version is available at ACM Digital Library: \href{https://dl.acm.org/doi/10.1145/3701716.3715243}{DOI 10.1145/3701716.3715243}.}

The success of businesses is highly dependent on their ability to convert consumers into loyal customers. The value proposition of a customer is considered the primary determinant in this conversion process, and establishing intrinsic product worth at a specified value point is essential. While affordability remains a significant factor for online shoppers, brands are transitioning away from broad-scale promotions and campaigns that reduce the perceived brand worth. Implementing such widespread marketing campaigns can inadvertently erode brand equity and diminish the probability of achieving the desired return on marketing investment. Existing solutions have fallen short in addressing the twin challenges of finding the optimal balance between the value proposition and the long-term brand value. At the same time, existing dynamic economic algorithms often misidentify highly engaging customers as ideal campaign targets, leading to suboptimal engagement and conversion rates, thereby diminishing customer loyalty.

This article introduces a two-stage multi-model architecture that employs Self-Paced Loss to enhance customer categorization. The first layer is a Multi-Class Neural Network that differentiates high engaging customers stimulated by a campaign, high-engagement customers not influenced by campaign nudges, and low-engagement customers. The Binary Label Correction Model forms the second layer, further refining the classification of highly engaged customers by differentiating between those responding to a campaign and those engaging organically. The actual customer’s intent can be determined using a missing label framework. This level gathers data from customers displaying variable value proposition behavior, rectifying their true state labels during training for more efficient customer segmentation. By distinguishing customer engagement intent (prompted versus organic), the suggested solution allows businesses to enhance their marketing campaign strategies and target prompted engaged segment, reducing exposure rates while boosting conversion rates.  When testing this solution using an A/B test framework, we noticed an increase of more than 100 basis points in the success metric.

\end{abstract}

%%
%% The code below is generated by the tool at http://dl.acm.org/ccs.cfm.
%% Please copy and paste the code instead of the example below.
%%


%%
%% Keywords. The author(s) should pick words that accurately describe
%% the work being presented. Separate the keywords with commas.

%%
%% This command processes the author and affiliation and title
%% information and builds the first part of the formatted document.
\maketitle

\section{Introduction}

In highly competitive business landscape nowadays, understanding and catering to customer behavior and preferences is paramount for businesses to maintain and expand market share \cite{Akbar2024}. Traditional customer segmentation methods, which often rely on static demographic information, are increasingly being replaced by more personalized approaches \cite{Wedel2016, Kumar2018}. The advent of big data and advancements in machine learning have enabled businesses to analyze vast amounts of data in real-time \cite{Perumalsamy2022}, providing deeper insights into customer needs and behaviors \cite{Gandomi2015, Wedel2016}. This shift towards dynamic customer segmentation allows for more personalized marketing strategies, improved customer satisfaction, and increased profitability \cite{Rust2004, Vesanen2007}.

While these personalized marketing campaigns and offers are primary influences for online buyers \cite{monroe2002}, in today's competitive market, simply launching site-wide campaigns and offers may not be enough to attract and retain customers. Mental accounting perspective suggests that customers evaluate products based on their perceived value \cite{thaler85}. In the context of personalized marketing campaigns and offers, understanding the customer's perceived value of a product can help retailers tailor their pricing strategies to individual customers. 

In the context of current economic uncertainties, a significant number of consumers have developed a more discerning approach to seeking value for their money \cite{Sheth2020}. This shift in consumer behavior has led to a keener focus on gaining maximum value from each dollar spent. At one end of the spectrum, there are certain dynamic individuals and consumers from Generation Z who have exhibited a noticeable change in their product value preferences \cite{Djafarova2020}. They are more inclined to purchase items in higher-priced categories and demonstrate a readiness to pay the full price or even more for certain products \cite{wilson2019}. These customers do not place a high reliance on marketing campaigns or special offers to drive their purchasing decisions.

On the other end of the spectrum, there is another faction of consumers who tend to respond more to marketing campaigns and offers \cite{Gupta2003}. This group of consumers often require some form of prompt from a campaign before they decide to make a purchase. However, this group of consumers can be transformed into loyal customers if they are presented with competitive pricing and promotions that align with their budgetary constraints. By acknowledging and understanding the diverse needs and preferences of these two distinct consumer groups, retailers can develop targeted marketing strategies and flexible pricing models to cater to them \cite{Potluri2024}. This approach can enable retailers to appeal to a wider range of consumers, thereby potentially enhancing their sales figures and fostering greater customer loyalty.

Understanding the dynamics between consumers who respond to marketing campaigns and those who transact organically is essential for businesses aiming to craft effective strategies \cite{Jeffrey2005, Moe2004}. Each group exhibits distinct behaviors and characteristics, impacting customer acquisition costs, loyalty, and long-term value \cite{Majka2024}. Thus, the value proposition of a customer plays a predominant role in the success of these marketing campaigns. The value proposition of a customer represents their perceived worth to a business, balancing the potential profitability they bring against the cost of engaging them. It encompasses their likelihood to respond to campaigns, transact organically, and contribute to long-term brand equity \cite{Bleier2015}.

\textbf{Consumers Responding to Marketing Campaigns}: Consumers who respond to marketing campaigns are primarily influenced by targeted promotional efforts such as advertisements, email campaigns, social media ads, and special offers \cite{Ashley2015}. These consumers are often drawn to businesses due to compelling discounts, limited-time deals, or eye-catching advertisements. They exhibit high price sensitivity and are more likely to make impulse purchases driven by immediate incentives \cite{Chen2012}.

The engagement level of these consumers with the brand is typically higher during the campaign period, as they interact with various marketing channels \cite{Laroche2014}. However, their loyalty may be short-lived, often contingent upon the continuation of promotions \cite{Dastane2021}. This group tends to have a higher Customer Acquisition Cost (CAC) since businesses need to invest in marketing efforts to attract them \cite{Majka2024, Zhang2018}. While marketing campaigns can provide a significant sales boost in the short term, businesses must implement effective retention strategies to convert these consumers into repeat customers. Importantly, marketing campaigns generate valuable data on consumer preferences and behaviors, which can be leveraged to refine future marketing efforts \cite{Kar2023}.

\textbf{Consumers Transacting Organically}: In contrast, consumers who transact organically discover and purchase products without the direct influence of marketing campaigns. Their decision-making process is driven by personal needs, brand reputation, word-of-mouth recommendations, and organic search results \cite{Reichheld2000}. These consumers often exhibit strong brand loyalty and trust, having conducted thorough research and read reviews before making a purchase.

Organic consumers have a lower CAC since they are not acquired through paid marketing efforts \cite{Ramos2023}. They tend to provide a more stable and predictable revenue stream due to their repeat purchases and higher Customer Lifetime Value (CLV) \cite{Singh2010}. This group’s purchasing decisions are influenced by the brand's reputation and the quality of the customer experience, making it crucial for businesses to maintain a positive brand image and deliver exceptional service\cite{Fawad2014}. Investments in search engine optimization and content marketing are pivotal in enhancing organic reach and attracting these consumers \cite{Shad2024}.

\textbf{Comparing and Balancing Both Groups}: Marketing campaign responders offer the advantage of a quick sales boost and measurable return on investment (ROI), but they require continuous promotional efforts to maintain engagement and loyalty \cite{Gupta2011}. On the other hand, organic consumers provide sustainable revenue and higher loyalty, relying on the strength of the brand and organic discovery channels.

To achieve a balanced approach, businesses should integrate marketing efforts to attract new customers while leveraging organic strategies \cite{Sahar2018} to retain them \cite{Lemon2016}. Personalization based on consumer data can enhance engagement and long-term loyalty \cite{Potter2024}. Emphasizing an exceptional customer experience is critical in converting campaign responders into loyal customers. The continuous analysis of data from both groups allows businesses to optimize their strategies and improve overall performance \cite{Krishnamoorthi2024}.

By understanding and leveraging the strengths of both consumer groups, businesses can develop more comprehensive and effective marketing strategies, ensuring both immediate sales boosts and long-term growth.

\section{Related Work}
In marketing research, identifying target audiences with precise value propositions is critical to personalizing marketing campaigns and optimizing their effectiveness \cite{Bleier2015}. The application of machine learning models to personalized marketing is a rapidly evolving field, showing significant potential to revolutionize marketing strategies for e-commerce companies. Existing works primarily focus on predictive modeling of customer behaviors using supervised approaches such as tree-based models \cite{GUELMAN201524, kim2001} and neural networks \cite{CHEN2021102573, HUANG202082}. Meanwhile, unsupervised methods, including k-means clustering \cite{gayathri2021, PAPAMICHAIL20071400} and its integration with traditional RFM (Recency, Frequency, Monetary) marketing techniques \cite{sarkar2024optimizing}, have also gained considerable traction.

Traditional marketing approaches like RFM have been widely adopted due to their simplicity and effectiveness in targeting high-value transactional customers. However, these methods often rely on generalized cohort-level analyses based purely on high transactional metrics (i.e., high recency, frequency, and monetary value) without adequately personalizing campaigns or pricing for individual customers. Such approaches fail to account for nuanced customer behaviors, particularly the varying sensitivities to value propositions and pricing strategies.

To address these limitations, companies like Amazon have implemented advanced dynamic pricing methods such as the "Intelligent Pricing" initiative \cite{Cooprider2023}. This approach uses a two-pronged strategy: analyzing both historical and real-time market data to understand evolving consumer price sensitivities, and recommending personalized price adjustments at critical decision-making moments. While effective in some scenarios, these methods still lack granularity, particularly in distinguishing between price-sensitive customers and those willing to pay full price organically. Consequently, dynamic pricing systems often result in blanket price adjustments that overlook customer-specific behaviors, leading to suboptimal outcomes.

Another prominent technique in the domain of campaign optimization is uplift modeling \cite{baier2022profit, rzepakowski2012uplift}, which measures the incremental impact of marketing actions on individual behaviors. Although uplift modeling is effective in evaluating campaign effectiveness, its reliance on high-quality experimental data with well-designed control and treatment groups presents a significant challenge in real-world scenarios. The resulting data sparsity and missing labels, particularly for customers who have never been subjected to campaigns, further limit its applicability.

Our proposed method addresses several of these limitations. Unlike traditional RFM techniques, it incorporates value proposition features to create a more holistic customer segmentation framework. By distinguishing between price-sensitive and non-price-sensitive users, our approach avoids the pitfalls of generalized dynamic pricing systems, ensuring personalized interventions for each customer segment. Additionally, our method overcomes the missing label issue inherent in uplift modeling by employing robust label correction techniques within a two-stage model architecture. This enables precise targeting of both highly transactional and value-driven customers, significantly enhancing campaign effectiveness and customer engagement.

%understanding customers' price sensitivities is crucial for personalizing campaigns and identifying target audiences. The application of machine learning models to predict price sensitivity is a rapidly evolving field with significant potential to transform marketing strategies for e-commerce companies. Classic supervised learning models such as have been widely used \cite{CHEN2021102573}

\section{Methodology}

\subsection{Latent Customer Segmentation}
Based on transaction history and engagement with marketing initiatives, we identify three customer segments:
\begin{itemize}
    \item Engaged segment consists of customers who have previously engaged with our marketing campaigns and have made a subsequent purchase.
    %: 1) (value propositition customers do a decision-driven transaction) without ads, they would turn to other PTs. 2) (mindless transaction) without ads, they stick to same PT
    \item Unengaged segment consists of customers who have not engaged with our marketing campaigns and have made purchases without any known marketing influence. 
    \item Inactive segment consists of customers who have not engaged with our marketing campaigns and have not made any recent purchases. %They are a group of customers who have not demonstrated any known purchase behavior or engagement with our marketing efforts.
\end{itemize}

Understanding these customer segments can help tailor our marketing strategies and increase the effectiveness of our campaigns. However, the segmentation only considers customer behaviors while overlooking their underlying intents, particularly for the engaged group which could have two different motivations. 
Some engaged customers are prompted, and due to lower perceived value, they could potentially shift their loyalty to competing products or brands in the absence of such campaigns. Conversely, other organically engaged customers, who have similar user pattern as the unengaged segment, might continue their purchases irrespective of any marketing influence.

To reduce marketing costs and boost marginal profits, Walmart and other e-commerce companies aim to focus on the first scenario, targeting customers who are undecided and can be nudged by marketing campaigns. 
This requires model to not only predict customer responses to campaigns precisely, but also comprehends the underlying intent, which is often not directly observable and results in a missing label issue.

\subsection{Overall Architecture}
\begin{figure}
    \centering
    \includegraphics[width=0.9\linewidth]{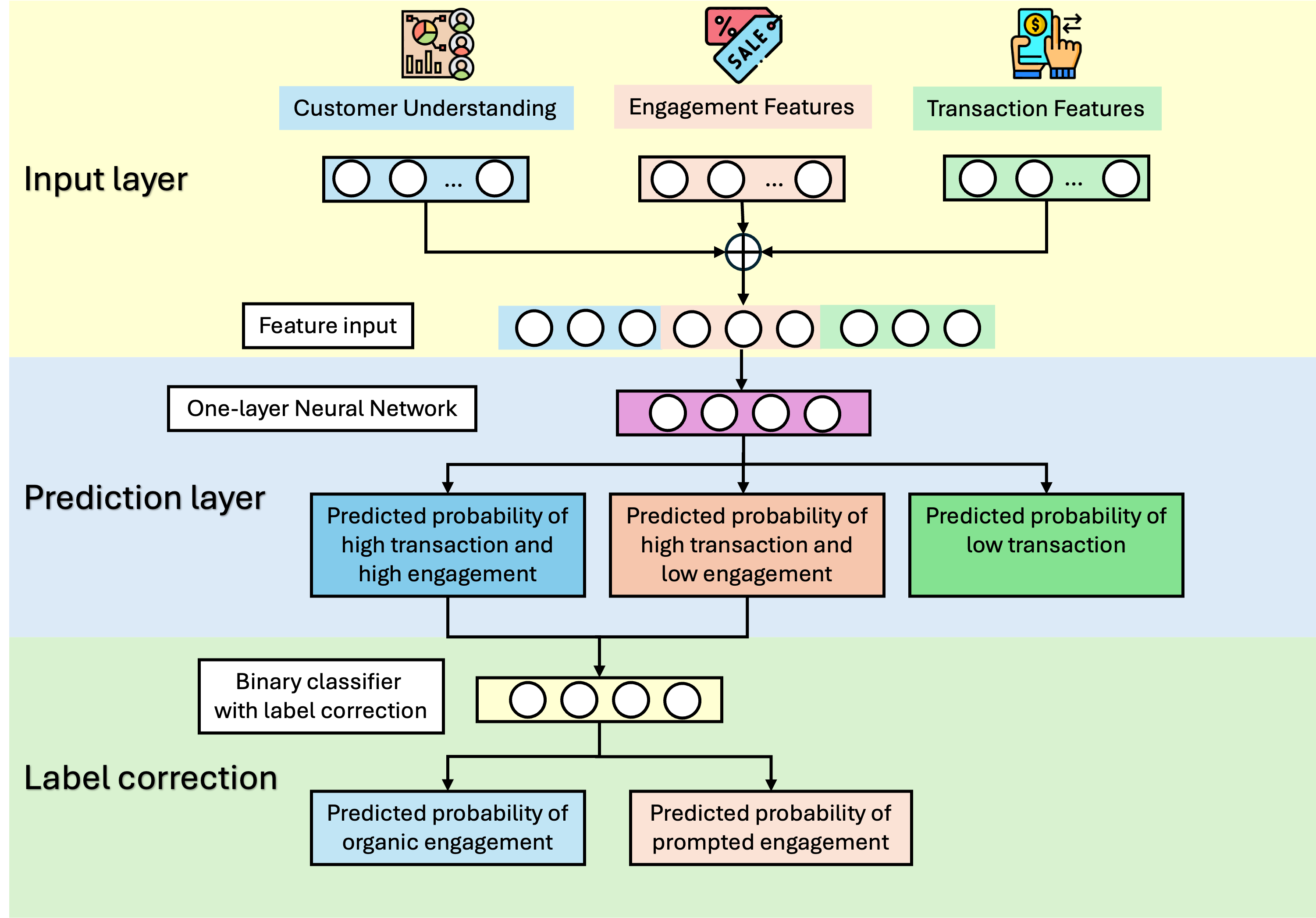}
    \caption{The proposed model architecture for predicting customer behaviors and intent, including label correction and incremental gain identification.}
    \label{diagram}
\end{figure}

To solve the missing label issue and capture customer intent behind their behaviors, we propose a two-stage architecture as shown in Figure \ref{diagram}. 
The proposed architecture predicts customer behaviors and excludes inactive customers in the first stage. In the second stage, the proposed architecture further predicts customer intent with label correction, and identifies the true incremental gains achievable through marketing promotions, i.e. customers who will respond but would not have made a purchase without exposure to campaigns. 

In the first stage, we train a single-layer neural network for multi-class classification, and three segments defined in Section 3.1 are labeled as three classes. Considering the imbalanced sample sizes of three classes, we adopt a weighted categorical cross entropy loss given by
\begin{equation}
    \mathcal{L}_{\rm CCE} = -\sum_{i=1}^{N} \sum_{c=1}^{C} w_c y_{i,c} \log(p_{i,c}), 
\end{equation}
where $N$ denotes the number of samples, $C$ denotes the number of classes, $w_c$ is the weight for class c, $y_{i,c}$ is a binary indicator (i.e. 0 or 1) if class c is the correct classification for the i-th sample, and $p_{i,c}$ is the predicted probability that the i-th sample is of class c. Here each class c is given a weight $w_c$, which adjusts the contribution of that class to the total loss and helps to address the class imbalance by giving more importance to underrepresented classes. 

For the i-th sample, the model predicts a probability vector $\hat{p}_i=(\hat{p}_{i,0}, \hat{p}_{i,1}, \hat{p}_{i,2})$, where each component represents the predicted probability of the sample customer belonging to the corresponding segment, respectively. Given the model prediction on customer behaviors, we can distinguish those customers with high predicted probabilities of belonging to the inactive segment, and focus on customers more likely to have engagement with marketing campaigns.

In the second stage, we aim to distinguish value proposition customers and predict if customers will make purchases without any marketing stimuli. But for the engaged group who have been exposed and engaged with marketing campaigns, whether they would have purchased the product without exposure is unknown. So we start with initial labels generated by customer behaviors and make label correction at further iterations. Initial labels are assigned as labeling the unengaged segment as the negative class that will make purchases without any marketing incentives, while labeling the engaged segment as the positive class. True positive labels are users who promptly engaged and motivated by marketing incentives, while true negative labels are organically engaged without marketing. In the initial labels, we could have mislabeled some customers in the engaged group who would have made a purchase even without campaigns. 

Thus for label correction, we need a robust loss function of binary classification, which can flip negative labels to positive if predicted probability is lower than a given threshold. So we start with a binary cross entropy (BCE) loss given by 
\begin{equation}
    \mathcal{L} = -\frac{1}{N} \sum_{i=1}^{N} \left[ y_i L_i^+ + (1 - y_i) L_i^- \right], 
\end{equation}
where $N$ denotes the number of users, $y_i$ denotes the true label for the i-th user, and 
$L_i^+$ and $L_i^-$ denote the positive and negative losses of the i-th user. The positive and negative losses are given by
\begin{equation}
    \begin{cases}
       & L_i^+ = \log p_i \\
       & L_i^-= \log (1-p_i)
    \end{cases},
\end{equation}
where $p_i$ denotes the predicted probability of the i-th user being in the positive class. 

To make the BCE loss adaptive to correct false positives, we want $L_i^+$ to fall back to $L_i^-$ when model prediction suggests the i-th user is highly likely to engage organically. Following \cite{zhang2021simplerobustlossdesign}, we adopt the self-paced loss
correction (SPLC) and change the positive loss to
\begin{equation}
 L_i^+=   \begin{cases}
       & \log p_i, \text{ if } p_i>\tau \\
       & \log (1-p_i), \text{ if } p_i\leq\tau
    \end{cases},
\end{equation}
where $\tau\in (0,1)$ denotes a threshold to identify whether it is a true positive or false positive based on the model prediction probability $p_i$. Specifically, for a given positive label, it would be corrected with low probability $p_i>\tau$. 

To make the robust loss function more flexible to incorporate business insights, we further define the threshold $\tau$ as a tunable parameter. Then finalized loss function is given by
\begin{equation}
    \begin{cases}
       
       L_i^+(\theta) =& \mathcal{I}(p_i(\theta)\leq\tau(\theta))\log (1-p_i(\theta)) \\
       &+ \mathcal{I}(p_i(\theta)>\tau(\theta))\log p_i(\theta) \\
       L_i^-(\theta) =& \log (1-p_i(\theta))
    \end{cases},
\end{equation}
where $\mathcal{I}$ denotes an indictor function and $\theta$ denotes parameters of binary classifier. 
For the i-th sample, the model generates a score as $p_i$, which is the predicted probability of the customer only converting if exposed to marketing campaigns. In the tailored marketing strategy, we tend to prioritize those customers with high scores due to marketing costs.

\begin{figure}
    \centering
    \includegraphics[width=0.75\linewidth]{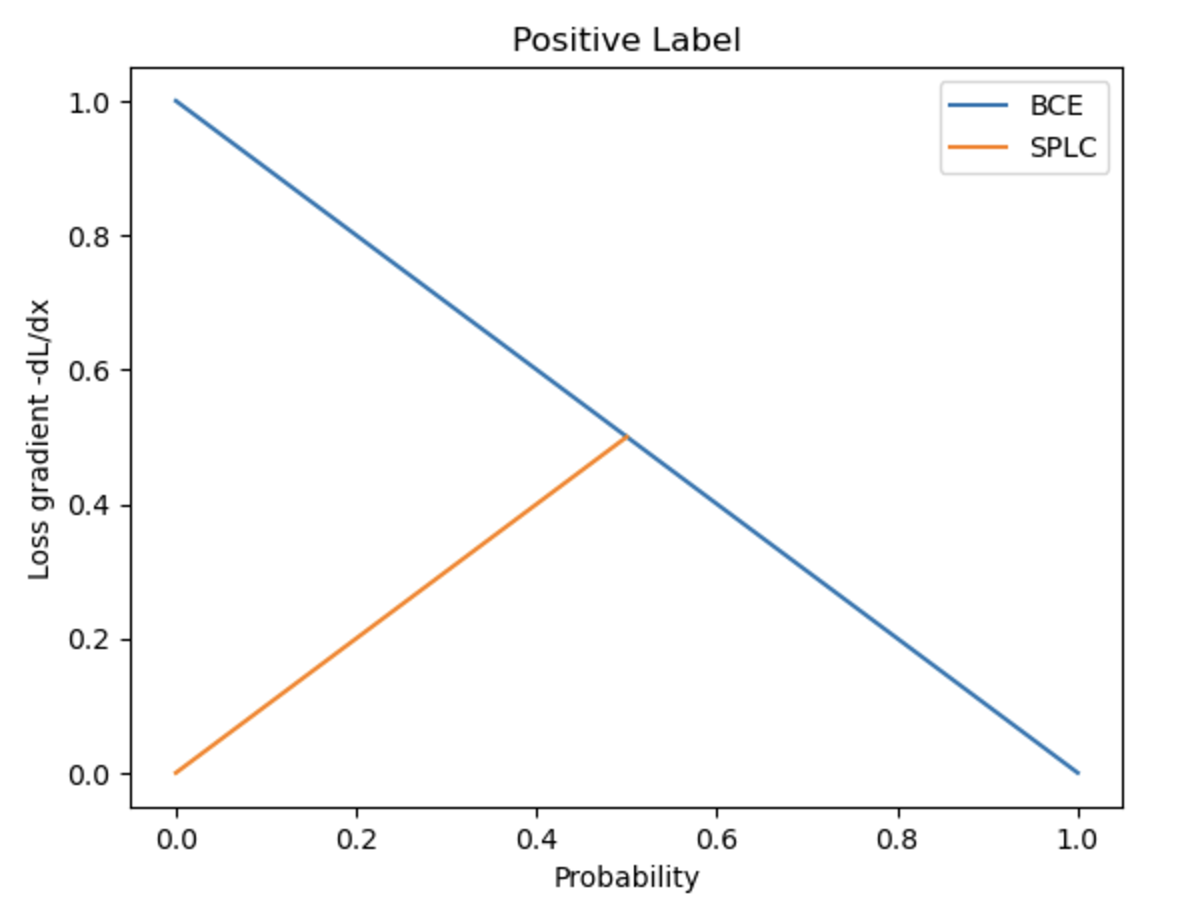}
    \caption{Loss gradient by probability, illustrating the relationship between predicted probability and loss gradient for SPLC.}
    \label{loss gradient}
\end{figure}

Figure \ref{loss gradient} demonstrates how the SPLC adapts to false positives with an example of $\tau(\theta)=0.5$. Here, the x axis is the predicted probability $p_i $, and the y axis is the loss gradient $-\frac{dL}{dx}$ where $x$ is the predict logit and $y=(1+e^{-x})^{-1}$. For a positive label, the smaller the probability, the more likely it is a false positive. Compared to the BCE, for which the loss increases as the predict probability decreases, the SPLC loss function assigns a lower weight to the false positives and reduces the effect of missing labels.

\section{Experiment}

In this section, we will discuss the offline model evaluation, and the online experiments and deployment in detail. We first introduce the dataset and the evaluation metrics used in this paper. Then, we illustrate the effectiveness of model by comparing to a strong baseline model widely used in online advertising. Finally, we present the successful online performance of the model on the Walmart promotion engine.

\subsection{Dataset}
We collected a sample of marketing response data and utilized a variety of features that help us gain insight into customer behavior and preference. User features were collected from user data before marketing started and labels were collected from user response after marketing initiatives were launched. 

User labels for the multi-class neural network at the first stage are generated based on user behaviors including engagement with campaign and transactions after that. For the binary label correction stage, true user labels are generated on users who were initially not included in marketing events but got exposure later, so we could identify if their engagement is prompted or organic.
 
User features are primarily divided into three categories: customer understanding, user engagement, and transactions. 
Customer understanding features include their family type, price affinity, fulfillment preferences, and personas. User engagement metrics include all impressions, views and click interactions on banners, campaigns, and splash pages. Data was collected from logs of all Walmart’s platform experiences: desktop web, mobile web, and IOS and Android native apps. Transaction features include the customer's transaction history with Walmart.

\subsection{Baseline and Metrics}
To illustrate that the optimal campaign audience are not highly engaging customers but high value proposition customers, we use a transaction-based user segmentation called RFM segment as the baseline model, compared to our proposed value proposition based user segment. RFM is a marketing technique used to quantitatively rank and group customers based on the recency, frequency and monetary total of their recent transactions. We selected the top 3 RFM segments of the most active transacting customer segments (i.e. champions, loyal customers and prospects) as our baseline model.

For comparison of classification models, we report the precision, recall, balanced accuracy and weighted F1-score of the
models as the metrics to evaluate their performance. The definitions
of these metrics are listed as follows:
\begin{itemize}
\item Precision measures the proportion of positive predictions that are actually correct.
\item Recall measures the proportion of true positive cases that the model correctly identifies.
\item Balanced accuracy is the average of sensitivity (recall for the positive class) and specificity (recall for the negative class).
\item Weighted F1-score is a weighted average of the F1 scores for each class, where the weights are proportional to the number of samples in each class.
\end{itemize}

\subsection{Offline Evaluation}

We evaluated the offline model performance compared to the top 3 RFM segments that are most active in transactions.

Table \ref{offline eval} shows the proposed model outperforms the baseline on the common evaluation metrics for a classification model such as precision, recall, balanced accuracy and weighted F1 score.

%It is important to highlight this is an imbalanced dataset in which the engaged group was underrepresented. This characteristic of the dataset can potentially introduce bias towards the majority class and challenge the model performance. Therefore, we used balanced accuracy instead of accuracy to avoid favoring the majority class.

%The threshold of 0.4 gives up a 0.724 balanced accuracy and 0.817 weighted F1 score. Furthermore, by stratifying the prediction scores into different thresholds, it becomes feasible to identify the optimal threshold that best aligns with our use cases. 

\begin{table*}
  \caption{Offline evaluation results.}
  \label{offline eval}
  \begin{tabular}{l l l l l }
    \toprule
     & precision & recall & Balanced accuracy & Weighted F1 score \\
    %threshold & precision & recall & Balanced accuracy & Weighted F1 score& True Positive Rate & False Positive Rate  \\
    \midrule
    Champion (RFM) & 0.371 & 0.433 & 0.624 & 0.746 \\
    Loyalist (RFM) & 0.280 & 0.207 & 0.537 & 0.718 \\
    Potential Loyalist (RFM) & 0.273 & 0.132 & 0.522 & 0.720 \\
    Model segment & \bf{0.448}  & \bf{0.600}  & \bf{0.724}  & \bf{0.817} \\
    
    %0.1  & 0.171  & 0.989  & 0.504  & 0.082  & 0.989  & 0.980  \\
    %0.2  & 0.285  & 0.874  & 0.712  & 0.653  & 0.874  & 0.449  \\
    %0.3  & 0.346  & 0.742  & 0.727  & 0.750  & 0.742  & 0.288  \\
    %0.4  & 0.448  & 0.600  & 0.724  & 0.817  & 0.600  & 0.152  \\
    %0.5  & 0.560  & 0.475  & 0.699  & 0.842  & 0.475  & 0.077  \\
    %0.6  & 0.665  & 0.388  & 0.674  & 0.847  & 0.388  & 0.040  \\
    %0.7  & 0.751  & 0.306  & 0.642  & 0.840  & 0.306  & 0.021  \\
    %0.8  & 0.790  & 0.202  & 0.595  & 0.817  & 0.202  & 0.011  \\
    %0.9  & 0.767  & 0.092  & 0.543  & 0.785  & 0.092  & 0.006  \\
    \bottomrule
  \end{tabular}
\end{table*}

\begin{figure}
    \centering
    \includegraphics[width=0.75\linewidth]{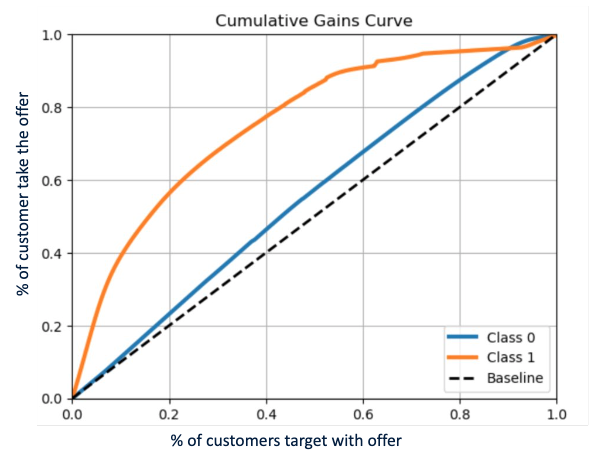}
    \caption{Cumulative gain curve. The curve illustrates how many customers to target and what returns to expect from a marketing campaign.}
    \label{CGC}
\end{figure}

Figure \ref{CGC} shows the cumulative gain curve to further illustrate that the proposed model aids in resource allocation by focusing on the most promising customers. X axis represents the percentage of customers that have been targeted, and y axis represents the cumulative percentage of actual promptly engaged customers that can be captured in target audience. Figure \ref{CGC} reveals that by targeting 40\% of model-selected customers, 78\% of the actual value proposition customers can be reached. That is, despite a reduction in the overall campaign exposure rate, we can achieve a higher conversion rate. The curve helps us make informed decisions regarding the number of customers required to be targeted for achieving marketing objectives.

%X axis: This represents the percentage of customers we target, starting from the highest-ranked (on the left) to the lowest-ranked(on the right)
%Y Axis: This represents the cumulative percentage of actual value proposition customers captured

\subsection{Online evaluation}

To assess the efficacy of our model in real-world scenarios, we conducted an online A/B testing experiment. The marketing campaign targets a considerable user base of 10.8 million, with 6.6 million users who have never been part of Walmart campaigns, and 4.2 million users who used to engage with campaigns but have been churned. This test was carefully structured to incorporate two concurrently running setups: a control group and a treatment group. 

To ensure balanced sample sizes and unbiased results, we randomly split the users into a control group and a treatment group in a stratified manner to make sure the population sizes of churned and never users are the same in the control and treatment group, respectively. For the treatment group, we ranked all the users based on their model scores and selected the top users to show the offers. For the control group, we random sampling a subset to get an equal amount of users. Both the control and the treatment group got identical exposure to the marketing campaign. 

 %5.4 million users in the control group, and 5.4 million in the treatment group. %This approach facilitated a comprehensive comparison of random and model-driven selection results, providing valuable insights into our model's effectiveness. 

%Furthermore, we targeted engaged and inactive customers for acquisition and evaluated the model performance in both segments. We separated our 10.8 million targets into 2 distinct groups. 6.6M were designated for targeting engaged customer, and the remaining 4.2M for targeting inactive customers. Following weeks of A/B testing, we gathered and analyzed the results, which are presented below. 

\begin{table*}
  \caption{Online evaluation results for new customers.}
  \label{never user}
  \begin{tabular}{l l l l}
    \toprule
    group & User count  & \%Site visit  & Conversion rate  \\
    \midrule
    Random Sample & 3.3M  & 79\%  & 0.17\%  \\
    Model Driven & 3.3M  & 96\%  & 0.34\%  \\
    Lift &  & \bf{21.52\%}  & \bf{99.85\%}  \\
    \bottomrule
  \end{tabular}
\end{table*}

\begin{table*}
  \caption{Online evaluation results for churned customers.}
  \label{churn user}
  \begin{tabular}{l l l l}
    \toprule
    & User count  & \%Site visit  & Conversion rate  \\
    \midrule
    Random Sample  & 2.1M  & 87\%  & 0.68\%  \\
    Model Driven  & 2.1M  & 90\%  & 0.76\%  \\
    Lift  &  & \bf{3.45\%}  & \bf{10.78\%} \\
    \bottomrule
\end{tabular}
\end{table*}

Table \ref{never user} and Table \ref{churn user} were obtained by comparing the results of random group and model group. On both user groups, there was a statistically significant increase in conversion rates. Specifically, there was a remarkable 99.85\% uplift among never customers, and a 10.78\% lift in previously churned customers. This data demonstrates the effectiveness of our model in targeting potential customers, resulting in a significant improvement in their overall engagement and interaction with the brand. 

\subsection{Online Deployment}
The proposed model has been deployed to support campaign targeting by emails and promotion engine for real-time user advertising since April 2024. 

\begin{figure}
    \centering
    \includegraphics[width=\linewidth]{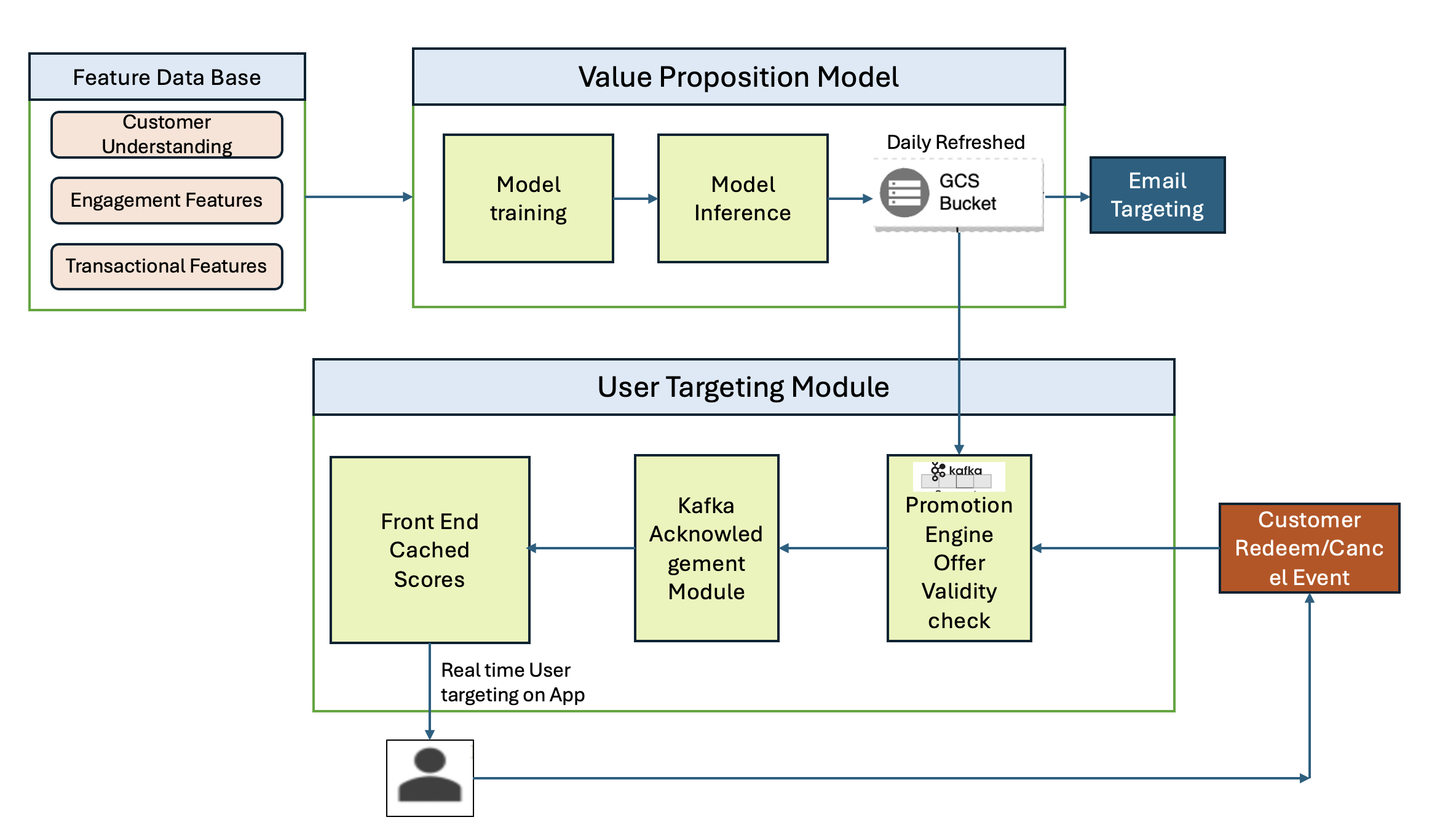}
    \caption{Overall framework of system deployment.}
    \label{deployment}
\end{figure}

Figure \ref{deployment} illustrates the role of the value proposition model in the Walmart ecosystem. The user features are collected and saved in the back end, and passed for model inference on a daily basis. The model refreshes user scores in a batched manner and updates in cloud-based storage. For email targeting, the optimal user list is directly retrieved from Google cloud storage. For online advertising triggered by real-time site visit, the model-driven user list will be checked by promotion engine for offer validity and sent to front end via Kafka acknowledge module. Customer response data keeps being feed back to promotion engine for logging and model re-training purpose.

%\begin{table*}
%  \caption{Model comparison}
%  \label{model comparison}
%\begin{tabular}{l l l l l}
    %\toprule
    %Metric  & Model Driven  & Champion  & Loyalists  & %Potential Loyalist  \\
    %\midrule
    %Balanced accuracy  & 0.73  & 0.63  & 0.53  & 0.52  \\
    %Weighted F1 score& 0.81  & 0.72  & 0.67  & 0.68  \\
    %precision  & 0.45  & 0.33  & 0.23  & 0.23  \\
    %recall  & 0.6  & 0.44  & 0.2  & 0.13  \\
    %\bottomrule

%\end{tabular}
%\end{table*}

%Table \ref{model comparison} listed the model performance comparison between . Our model outperformed the top RFM segments in all metrics, including balanced accuracy, weighted F1 score, precision and recall. Those metrics indicates that our model can better target the audience and captured those who are true value proposition customers. This shows that our model is able to generate appropriate personalized recommendations based on users’ shopping habits and transaction history and potentially acquire new customers given the right discount. The RFM method, in contrast, only targets high transaction customers, where they already have a strong connection and loyalty to Walmart brand. Without providing them with extra incentive, they will still react to the products. 

\section{Conclusion}

The paper introduced a two-stage multi-model architecture that utilizes Self-Paced Loss to improve customer categorization, addressing the issue of striking a balance between value proposition and long-term brand value. This system allows for a more efficient customer segmentation by discerning between customers who engage due to marketing campaigns and those who engage naturally. 

The best trained model was deployed to our production environment and evaluated in an online A/B test experiment. Based on model predictions, er achieved a nearly 100\% lift on conversion rate compared to the baseline model. Therefore, the proposed solution presents a novel approach to enhancing marketing strategies, reducing exposure rates, and increasing conversion rates. This paper highlights the need for dynamic customer engagement, paving the way for more complex, customer-centric strategies in the future. 
\vfill\eject

\printbibliography[notkeyword=move] 
\vfill\eject
\printbibliography[heading=none,keyword=move]

\end{document}